\documentclass[aps,prb,twocolumn,showpacs,superscriptaddress]{revtex4-1}
\usepackage{amsmath, array}
\usepackage{graphicx}
\usepackage{hyperref} 

\newcommand{\exponential}[1]{\mathrm{e}^{#1}}

\begin{document}

\title{On the equivalence of models with similar low-energy quasiparticles}

\author{Mirko M. M\" oller} \affiliation{\!Department \!of \!Physics and
  Astronomy, \!University of\!  British Columbia, \!Vancouver, British
  \!Columbia,\! Canada,\! V6T \!1Z1} \author{Mona Berciu}
\affiliation{\!Department \!of \!Physics and Astronomy, \!University
  of\!  British Columbia, \!Vancouver, British \!Columbia,\! Canada,\!
  V6T \!1Z1} \affiliation{\!Quantum Matter \!Institute, \!University
  of British Columbia, \!Vancouver, British \!Columbia, \!Canada,
  \!V6T \!1Z4}

\begin{abstract}
We use a Metropolis algorithm to calculate the finite temperature
spectral weight of three related models that have identical
quasiparticles at $T=0$, if the exchange favors the appearance of a
ferromagnetic background. The low-energy behavior of two of the models
remains equivalent at finite temperature, however that of the third
does not because its low-energy behavior is controlled by rare events
due to thermal fluctuations, which transfer spectral weight well below
the $T=0$ quasiparticle peaks and generate a pseudogap-like
phenomenology. Our results demonstrate that having $T=0$ spectra with
similar quasiparticles is not a sufficient condition to ensure that
two models are equivalent, {\em i.e.} that their low-energy properties
are similar. We also argue that the pseudogap-like phenomenology is
quite generic for models of $t$-$J$ type, appearing in any dimension
and for carriers injected into both ferromagnetic and antiferromagnetic backgrounds.
\end{abstract}
\date{\today}

 \maketitle

\section{Introduction}
All physics knowledge is built on the study of models. Formulating a
model for the system of interest is thus a key step in any project. Of
course, ``all models are wrong, but some of them are useful''
\cite{best}. This is because ideally, a model incorporates all
relevant physics of the studied system so that its solution is useful
to gain intuition and knowledge regarding some properties of
interest. At the same time, models discard details assumed to be
irrelevant for these properties. Even though this makes them
``wrong'', it is a necessary and even desirable step if the solution
is to not be impossibly complicated.

How to decide where lies the separation line between relevant and
irrelevant aspects for a given system and set of properties of
interest, is still an art.  A general guiding principle, based on
perturbation theory, is that high-energy states can be discarded
(integrated out) if one is interested in low-energy
properties. Consequently, it is assumed that models with identical
low-energy spectra provide equivalent descriptions of a system, and
therefore the simplest of these models can be safely used.

A prominent example is the modeling of cuprates. It is widely believed
that the Emery model \cite{Emery} can be replaced by the simpler
$t$-$J$ model to study their low-energy physics \cite{Lee,
  Dagotto}. The justification was provided by Zhang and Rice
\cite{Zhang-Rice} who argued that the low-energy states of the Emery
model are singlets formed between the spin of a doping hole hosted on
the four oxygens surrounding a copper and the spin of that copper, and
that the resulting quasiparticle is described accurately by the
$t$-$J$ model \cite{Anderson}.  Whether this is true is still being
debated \cite{Bayo,Hadi1,Hadi2}.

In this article we show that by itself, the condition that two models
have the same low-energy spectrum is not sufficient to guarantee that
they describe similar low-energy properties, despite widespread belief
to the contrary. Indeed, we identify three models that have identical
$T=0$ quasiparticles yet have very different behavior at {\em any}
temperature $T\ne 0$. The qualitative differences are
due to rare events controlled by thermal fluctuations, which lead to
a pseudogap-type of phenomenology. 

While our argument takes the form of a ``proof by counterexample'', we
also provide arguments that our findings are not merely an
``accident'' caused by our specific choice of models, but are more
general in nature. Specifically, we comment on its validity in
arbitrary dimensions and also for other types of magnetic coupling
which differ from the examples that are our main focus.

The remainder of this article is organized as follows: we introduce
the models in Section \ref{sec:Models}, and discuss our method of
solution in Section \ref{sec:Method}. The main results, which are for
a particle injected into a ferromagnetic (FM) background, are presented
in Section \ref{sec:FM_results}, while Section \ref{sec:AFM_results}
contains some results for an antiferromagnetic (AFM) background, which
further substantiate our claims. Short conclusions are presented in
Section \ref{sec:conclusions}.

\section{Models}
\label{sec:Models}
\begin{figure}[b]
  \centering 
  \includegraphics[width=\columnwidth]{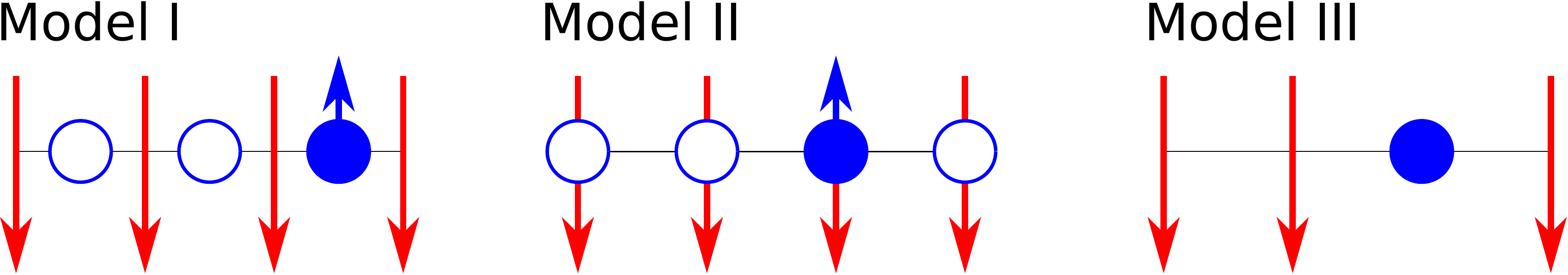}
  \caption{(color online) Sketch of the three models. Large, red
    arrows represent the local magnetic moments, empty (filled) blue
    circles represent empty (filled) carrier sites. For Models I and
    II the carrier spin is represented by a blue arrow, for Model III
    the carrier is a spinless ``hole'' in the Ising chain.}
  \label{fig:models}
\end{figure}

Because we are interested in the quasiparticle spectrum, from now on
we consider only the single carrier sector of the Fock space. To be
specific, we take the carrier to be an electron added into an
otherwise empty band; the solution is mapped onto that for removing an
electron from a full band by changing the energy $\omega \rightarrow
-\omega$.

The models of interest are sketched in Fig. \ref{fig:models}. They
describe the interaction of the carrier with a background of local
moments, and as such bear some similarity to those used in the
Zhang-Rice mapping mentioned above. Model I is the parent two-band
model, from which Models II and III are derived as increasingly
simpler effective models. In Model I, one band hosts the spin$-{1\over
  2}$ magnetic moments and a second band, located on a different
sublattice, hosts the carrier. Model II is also a two-band model, but
the carrier and local moments are located on the same sites. One can
think of the states occupied by the carrier in this model as being
local linear combinations of the carrier states in Model I, each
centered at a spin site. In Model III, the carrier is locked into a
singlet with its lattice spin, forming a ``spinless carrier''
analogous to the Zhang-Rice singlet.

There are also significant differences between our models and the
Zhang-Rice mapping: (i) we restrict ourselves to one dimension as this
suffices to prove our claim. However, some comments on the extension
of our results to higher dimensions can be found below; (ii) We
concentrate on the case of a ferromagnetic (FM) background because for
models with an antiferromagnetic (AFM) background the $T=0$ spectra
are not identical. However, we also present some AFM results later on,
to demonstrate that some of the features we discuss here are generic,
not FM-specific; and (iii) all spin exchanges are Ising-like, {\em
  i.e.} no spin flipping is allowed. The latter constraint allows us
to find numerically the exact solutions using a Metropolis algorithm
\cite{Ising-finite-T}, to uncover a surprising finite-$T$ behavior for
Model III.

In all three cases, the interactions between the local moments are
described by the Ising Hamiltonian:
\begin{align}
  \hat{H}_S = - J \sum_{i} \hat{\sigma}_{i+\delta}
  \hat{\sigma}_{i+1+\delta} -h \sum_i \hat{\sigma}_{i + \delta},
\end{align}
where $\delta=1/2$ for Model I and $\delta=0$ for Models II and III,
and $\hat{\sigma}_{i+\delta}$ is the Ising operator for the local
magnetic moment located at $R_{i+\delta} = i+\delta$ (we set
$a=1$). Its eigenvalues are $\sigma_i = \pm 1$. For $J>0$ the ground
state of $\hat{H}_S$ is FM, and it is AFM for $J<0$. In the case of
FM coupling, the external magnetic field $h$ can be used to favor
energetically one of the two possible FM ground states of the $h=0$
case.

For Models I and II, the kinetic energy of the carrier is described by
a nearest-neighbor hopping Hamiltonian:
\begin{align}
  \hat{T} = -t \sum_{i, \sigma} c_{i, \sigma}^\dagger c_{i+1, \sigma}
  + \text{h.c.}= \sum_{k,\sigma} \epsilon(k) c_{k,\sigma}^\dagger
  c_{k,\sigma}
\end{align}
where $c_{i, \sigma}^\dagger$ is the creation operator for a
spin-$\sigma$ carrier at site $R_i$ and $c_{k, \sigma}^\dagger = 
1/\sqrt{N}\sum_i \exponential{i k R_i} c_{i, \sigma}^\dagger$ are
states with momentum $k \in (-\pi,\pi)$ and eigenenergy $\epsilon(k) =
-2 t \cos k$ (the lattice constant is set to $a=1$). The interaction between the carrier and the local moments is
an AFM Ising exchange:
\begin{align}
  \hat{H}_{\text{ex}}^{(\text{I,II})} = \frac{J_0}{2} \sum_{i, \sigma}
  \sigma c_{i,\sigma}^\dagger c_{i, \sigma} (\hat{\sigma}_{i-\delta} +
  \hat{\sigma}_{i+\delta}).
\end{align}
Note that flipping the sign of the carrier spin corresponds to letting
$J_0 \rightarrow - J_0$, so we can assume without loss of generality
that the carrier has spin-up and suppress the spin index. The total
Hamiltonian for Models I and II is thus given by $\hat
H^{(\text{I,II})} = \hat H_S + \hat T + \hat
H_{\text{ex}}^{(\text{I,II})}$.

Model III is the FM ($J>0$) or AFM ($J<0$) Ising version of the
one-band $t$-$J$ model discussed extensively in the cuprate literature
\cite{Lee, Dagotto}.  The case of interest now has $N+1$ electrons in
the $N$ site system ($N\rightarrow \infty$), and double occupancy is
forbidden apart from the site where the additional carrier is located
and which can be viewed as hosting a ``spinless carrier'' whose motion
shuffles the otherwise frozen spins.  The Hamiltonian is $
\hat{H}^{(\text{III})} = \mathcal{P}\hat T\mathcal{P} +\hat H_S$,
where the operator $\mathcal{P}$\ projects out additional double
occupancy. It is important to note that in contrast to Models I and
II, here the spin-operators $\hat{\sigma}_i$ are related to the
electron creation/annihilation operators via $\hat{\sigma}_i =
\sum_{\sigma} \sigma c_{i,\sigma}^\dagger c_{i,\sigma}$.

\section{Method}
\label{sec:Method}
We calculate the finite-$T$ spectral weight $A(k,\omega) = -{1\over
\pi} \mbox{Im} G(k,\omega)$, where $G(k,\omega)$ is the one-carrier
propagator. If the carrier is injected in the magnetic background
equilibrated at temperature $T$, its real-time propagator is $G(k,
\tau) = -i \theta(\tau) \mbox{Tr}[ \hat{\rho}_S
c_{k\uparrow}(\tau)c_{k\uparrow}^\dagger(0)]$, where $\hat{\rho}_S =
\exp(-\beta \hat H_s)/Z$ is the density matrix of the undoped chain in
thermal equilibrium, $\beta = 1 /k_B T$ and $c_{k\sigma}(\tau) =
\exp(i \hat H \tau) c_{k\sigma} \exp(-i \hat H \tau)$ are the
operators in the Heisenberg representation (we set $\hbar=1$).

In frequency domain, the propagator becomes:
\begin{align}
  G(k,\omega) = \sum_{\{\sigma\}}\frac{\exponential{-\beta E^S_{\{
        \sigma \}}}}{Z} \langle \{ \sigma \}| c_{k\uparrow}
  \hat{G}(\omega + E^S_{\{ \sigma \}}) c_{k\uparrow}^\dagger | \{
  \sigma \} \rangle
\nonumber
\end{align}
The sum is over all configurations $\{ \sigma \}=(\sigma_1,
\dots,\sigma_N)$ of the Ising chain, with corresponding energies $\hat
H_S |\{ \sigma\}\rangle = E^S_{\{ \sigma \}}|\{ \sigma\}\rangle$, and
$Z= \sum_{\{\sigma\}}\exp(-\beta E^S_{\{ \sigma \}})$. The resolvent
is $\hat{G}(\omega) = [\omega - \hat{H} + i \eta]^{-1}$, where
$\eta\rightarrow 0^+$ ensures retardation. The shift by $E^S_{\{
\sigma \}}$ in the argument of the resolvent shows that the poles of
the propagator mark the {\em change} in the system's energy, {\em
i.e.} the difference between the eigenenergies of the system with the
carrier present, and those of the undoped states into which it was
injected. This reflects the well-known fact that electron addition
states have poles at energies $E_{N+1,\alpha}- E_{N,\beta}$
\cite{mahan}.

After Fourier transforming to real space and using the invariance to
translations of the thermally averaged system, we arrive at:
\begin{align} 
G(k,\omega) = \sum_n \exponential{i k R_n} \sum_{\{
\sigma \}}\frac{\exponential{-\beta E^S_{\{ \sigma\}}}}{Z}
g_{0,n}(\omega, \{ \sigma \}),
\label{eq:GF-FT} 
\end{align} 
where $ g_{0,n}(\omega, \{ \sigma \}) = \langle \{ \sigma \}|
c_{0,\uparrow} \hat G(\omega+E^S_{\{\sigma\}}) c_{n,\uparrow}^\dagger | \{ \sigma \}
\rangle$ is the Fourier transform of the amplitude of probability that
a state with configuration $\{ \sigma \}$\ and the carrier injected at
site $n$\ evolves into a state with the carrier injected at site
$0$. These real-space propagators are straightforward to calculate, as
they correspond to a single particle (consistent with our assumption of a  canonical ensemble with exactly one extra charge carrier in the system) moving in a frozen spin
background. We emphasize that this is true only because of the Ising nature of the exchange between the background spins. Heisenberg coupling, on the other hand, would lead to spin fluctuations that would significantly complicate matters. Below we present the calculation of these real-space propagators for Model III. For Model II, the solution is described in detail in Ref. \onlinecite{Ising-finite-T}, and the same approach,
with only minor modifications, applies to Model I.

It is convenient to introduce the following notation. When an extra
electron is injected at site $n$ of Model III it effectively removes
the spin at this site. The spin $\sigma_n$ will therefore be missing
from the set $\{ \sigma \}$ which describes the state of the
spin-chain before injection. Consequently we label the new state,
after injection, as $| \{ \sigma \} \setminus \sigma_n \rangle = |
\dots \sigma_{n-1} \circ \sigma_{n+1} \dots \rangle$, where $\circ$
denotes the effective ``hole'' created by the injection of the extra
electron. The ``hole'' can propagate along the chain and in doing so
reshuffles the spins. To capture the propagation of the ``hole'' we
introduce a new index $j$ corresponding to the number of sites that the
``hole'' has hopped to the left ($j<0$) or right ($j>0$). A general
state is therefore given by $|\{ \sigma \} \setminus \sigma_n, j
\rangle = | \dots \sigma_{n-1} \sigma_{n+1} \dots \sigma_{n+j} \circ 
 \sigma_{n+j+1} \rangle$. Note that this way of labelling states is
not unique. For instance, if $\sigma_0=\sigma_1= \dots = \sigma_n$, then $| \{\sigma\} \setminus \sigma_0,0\rangle = | \{\sigma\}
\setminus \sigma_n,-n\rangle$.

With this notation, the real-space propagators are
$g_{0,n}(\omega, \{ \sigma \}) = \langle \{ \sigma \} \setminus
\sigma_0, 0| \hat{G}(\omega + E^S_{\{ \sigma \}}) | \{ \sigma \}
\setminus \sigma_n,0 \rangle$. Their equations of motion (eom) are
obtained by splitting the Hamiltonian in two parts, $\hat{H} =
\hat{H}_0 + \hat{V}$, and repeatedly using Dyson's identity
$\hat{G}(\omega) = \hat{G}_0(\omega) + \hat{G}(\omega) \hat{V}
\hat{G}_0(\omega)$. Choosing $\hat{H}_0 = \hat{H}_S$ and suppressing the $\omega$ and $\{ \sigma \}$-dependence we obtain
\begin{align}
  g_{0,0} &= G_0(\omega + \Delta_0) [1 - t f_{0,1} - t f_{0,-1}], \\
  f_{0,n} &= -t G_0(\omega + \Delta_n)[f_{0,n+1} + f_{0,n-1}], 
\label{eq:EOM}
\end{align}
where $G_0(\omega)=(\omega + i\eta)^{-1}$, $\Delta_n = E^S_{\{\sigma\}} - E^S_{\{\sigma\} \setminus \sigma_0, n}$ and $f_{0,n} =\langle \{\sigma\} \setminus \sigma_0, 0| \hat{G}(\omega + E^S_{\{\sigma\}} ) | \{\sigma\} \setminus \sigma_0,n \rangle$. Note that $f_{0,0} = g_{0,0}$. The exact form of $\Delta_n$\ depends on the sign of $n$:
\begin{align}
  \Delta_0 &= -J \sigma_0(\sigma_{-1} + \sigma_1) \\ 
  \Delta_n &= \Delta_0 + J\sigma_{-1} \sigma_1 -J \sigma_n \sigma_{n+1},\ \text{for}\ n>0 \\ 
  \Delta_n &= \Delta_0 + J\sigma_{-1} \sigma_1 -J \sigma_n \sigma_{n-1},\ \text{for}\ n<0.
\end{align}

The eom (\ref{eq:EOM}) can be solved with the ansatz $f_{0,n} = A_n f_{0,n-1}$, for $n>0$\ and $f_{0,n} = B_{n} f_{0,n+1}$\ for $n<0$. Since the ``hole'' has a finite lifetime $\propto 1/\eta$\ and $f_{0,n}$\ measures the probability that the ``hole'' injected at site 0 moves to site $n$, one expects $f_{0,n} \rightarrow 0$\ for $n\rightarrow \infty$. We therefore introduce a sufficiently large cutoff $M_c$\ and require $A_{M_c}=0= B_{-M_c}$. It is then straightforward to obtain 
\begin{align}
  A_{n} &= \frac{-t}{\omega + \Delta_n +i \eta +t A_{n+1}}, &\\
  B_{n} &= \frac{-t}{\omega + \Delta_n +i \eta +t B_{n-1}}, &\\
  g_{0,0} &= \frac{1}{\omega + \Delta_0 +t B_{-1} + t A_1}, &\\
  f_{0,n} &= A_{n}\dots A_1 g_{0,0} & \text{for}\ n>0, \\
  f_{0,n} &= B_{n}\dots B_1 g_{0,0} & \text{for}\ n<0.
\end{align}

To calculate the $g_{0,n}$\ we make use of the fact that hopping reshuffles the spins. Therefore $g_{0,n} \neq 0$, only if $\sigma_0=\sigma_1=\dots =\sigma_n$. In that case, as mentioned above, the states $|\{\sigma\}\setminus \sigma_0,n \rangle$\ and $|\{\sigma\}\setminus \sigma_n,0 \rangle$\ are equal which means that $g_{0,n} = f_{0,n}$.

The thermal average in Eq. (\ref{eq:GF-FT}) is then calculated for the
infinite chain with a Metropolis algorithm which generates
configurations $\{ \sigma \}$ of the undoped chain. To summarize, our
method of solution consists of the following steps: {\em (i)} generate
a configuration $\{\sigma \}$ of the Ising chain using a Metropolis
algorithm; {\em (ii)} Calculate all the $g_{0,n}(\omega,\{ \sigma \})$
propagators for that specific configuration, and perform the sum over
$n$ in Eq. (\ref{eq:GF-FT}); {\em (iii)} repeat steps {\em (i)} and
{\em (ii)} until convergence is reached.  Full details of this
procedure can be found in Ref. \onlinecite{Ising-finite-T} for Model
II; the generalization to Models I and III is straightforward. For
Model III it is convenient to inject the carrier with an unpolarized
total spin, to ensure that a ``hole'' is always created. Since for
each configuration $\{\sigma \}$ there is a configuration $\{
\bar{\sigma} \}$ with all the spins flipped, injecting an unpolarized
carrier does not change the results, but merely speeds up the
numerics.

\section{Results}
\subsection{FM Results}
\label{sec:FM_results}
At $T=0$, the undoped Ising chain is in its FM ground state. The
quasiparticles of Models I and II have energy $\mp J_0 + \epsilon(k)$
if the carrier is injected with its spin
antiparallel/parallel to the background. Only the former case
can be meaningfully compared with Model III, which has a quasiparticle
of energy $2J + \epsilon(k)$ ($2J$ is the cost of removing two FM Ising
bonds). Thus, apart from trivial shifts, the three models have
identical quasiparticles, namely carriers free to move in the
otherwise FM background.

\begin{figure}[t]
  \centering
  \includegraphics[width=\columnwidth]{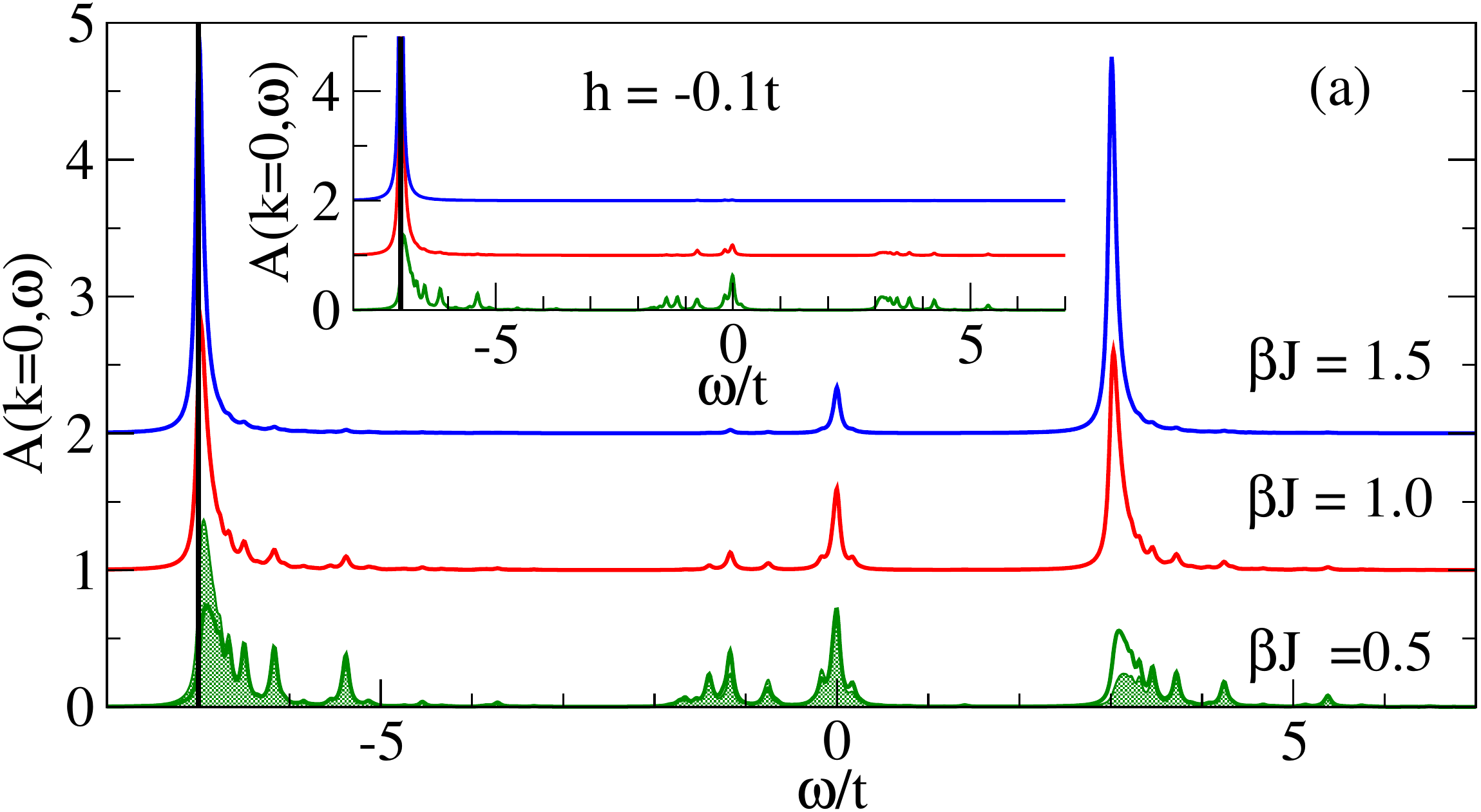}
  \includegraphics[width=\columnwidth]{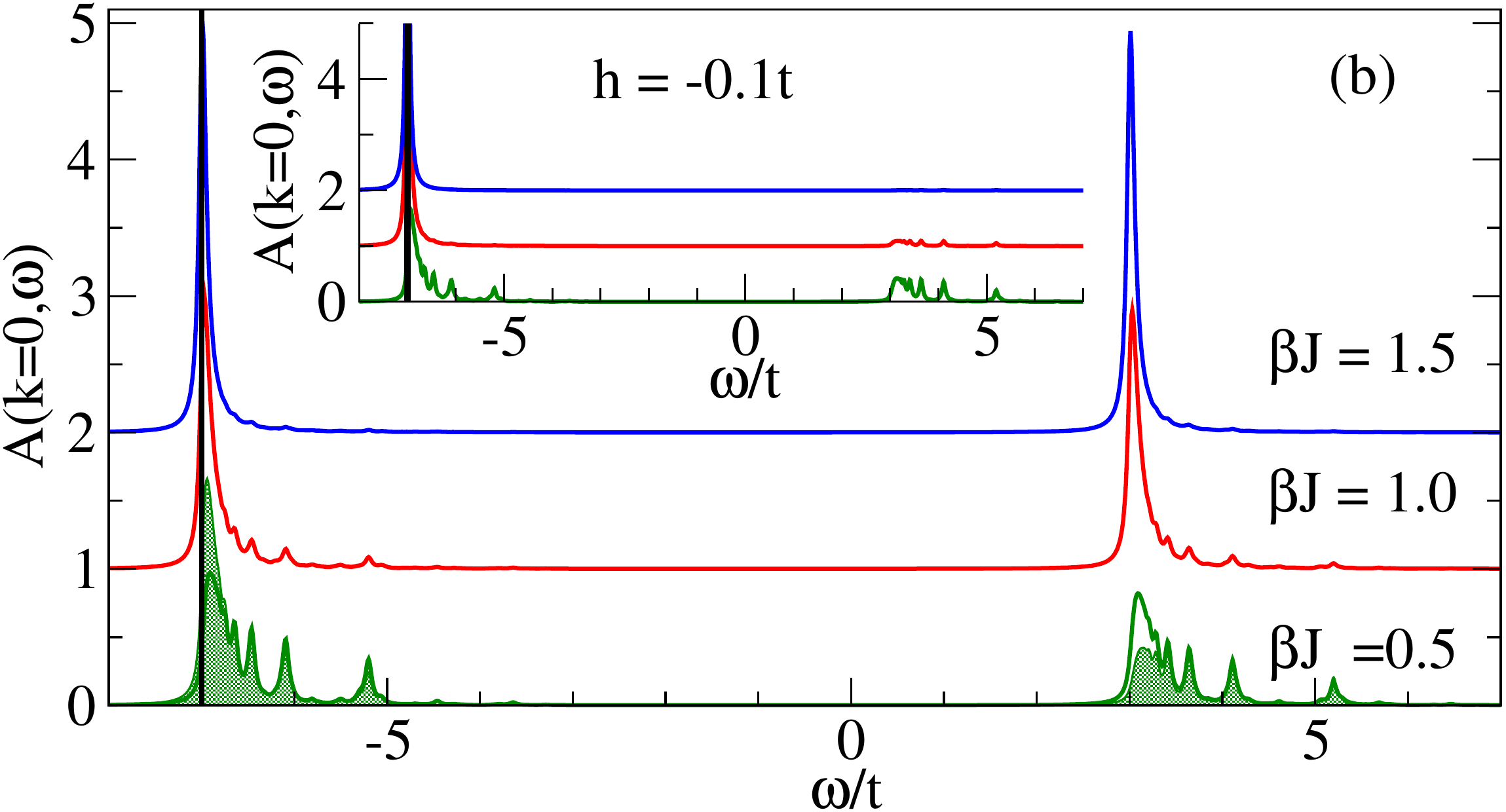}
  \includegraphics[width=\columnwidth]{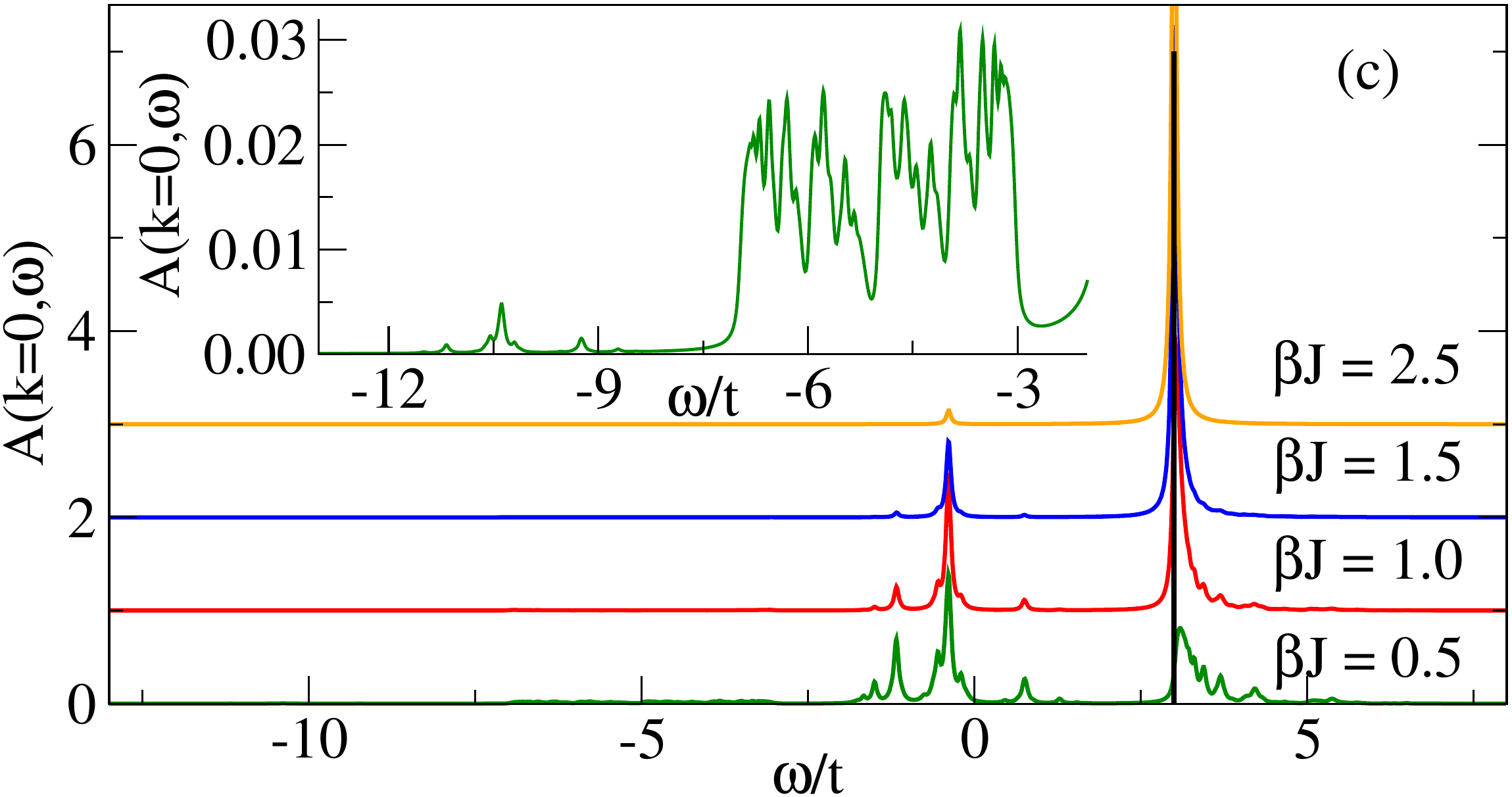}
  \caption{(color online) Spectral weight at $k=0$ for a FM background and three different
temperatures for (a) Model I with $J_0/t=5, J/t=0.5$; (b) Model II
with $J_0/t=5, J/t=0.5$; (c) Model III with $J/t=2.5$. Insets in
panels (a) and (b) show the spectral weight in the presence of a
magnetic field, while in (c) it shows the two continua appearing at
low energies, for $\beta J =0.5$. In all cases, the broadening is
$\eta/t =0.04$. The vertical lines show the energy of the $T=0$
quasiparticle peak.}
  \label{fig2}
\end{figure}

Finite-$T$ spectral weights $A(k=0,\omega)$ for the
different models are shown in Fig. \ref{fig2}.  We emphasize that
only the electron-addition part is discussed here. We do not consider
the electron-removal states, which lie at energies well below those of
the electron-addition states and must be identical for all three
models because in all cases, one of the electrons giving rise to the
magnetic moments is removed. We also emphasize that our calculation is
in a canonical ensemble. The chemical potential is not
fixed at $\omega =0$, 
as customary in grand canonical formulations, instead it can be
calculated as $\mu = \left({\partial F\over \partial N}\right)_T
\rightarrow \min_{\alpha,\beta}[E_{N+1,\alpha}-E_{N,\beta}]$ as
$T\rightarrow 0$. As pointed out above, here $\omega=0$ marks the
energy of the undoped Ising chain. 

For Models I and II, shown in panels (a) and (b), at the lowest
temperature one can see two peaks marking the contributions from
injection of the carrier into the two ground states of the Ising chain
(all spins up and all spins down, respectively). Indeed, these peaks
are located at $\pm J_0 - 2t$, the lower one of which is marked by the
vertical line.  Note that we chose a large $J_0$ value to keep
different features well separated and thus easier to identify. The
insets show the spectral weights for $h=-0.1t$, which at low-$T$
suppresses the contribution from the up-spin FM state so that only the
lower peak remains visible.

\begin{figure}[t]
  \centering 
  \includegraphics[width=\columnwidth]{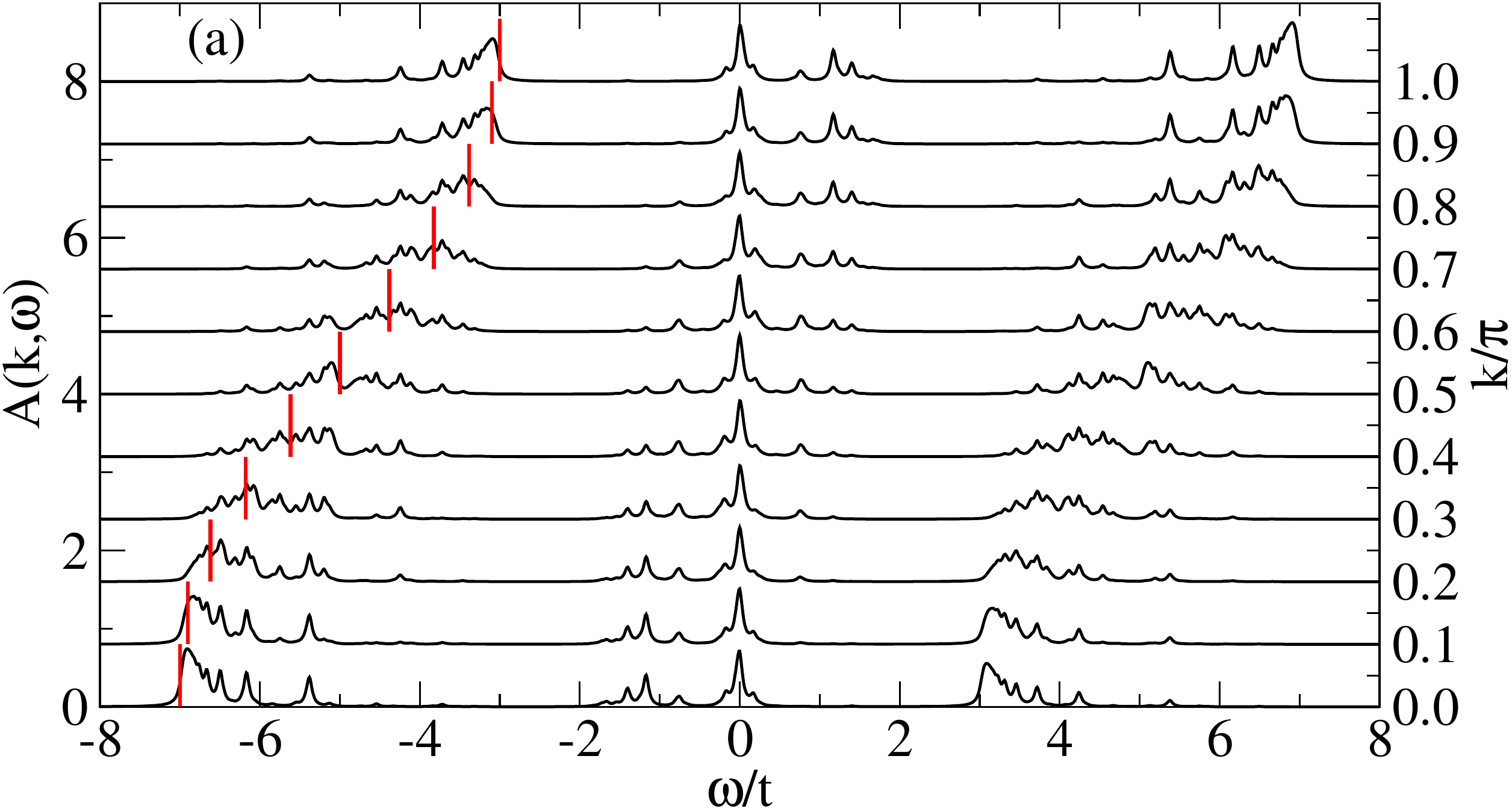}
  \includegraphics[width=\columnwidth]{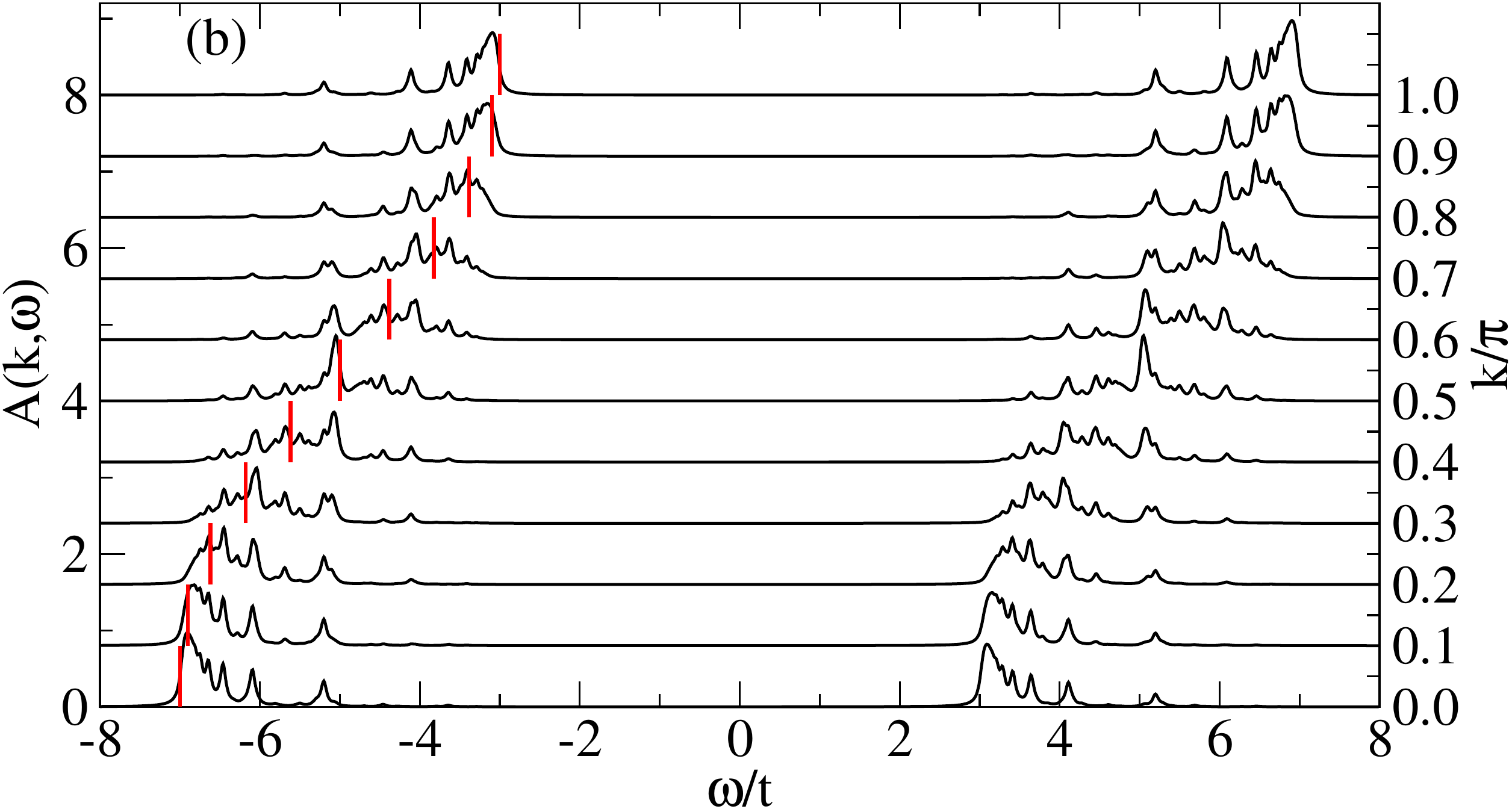}
  \includegraphics[width=\columnwidth]{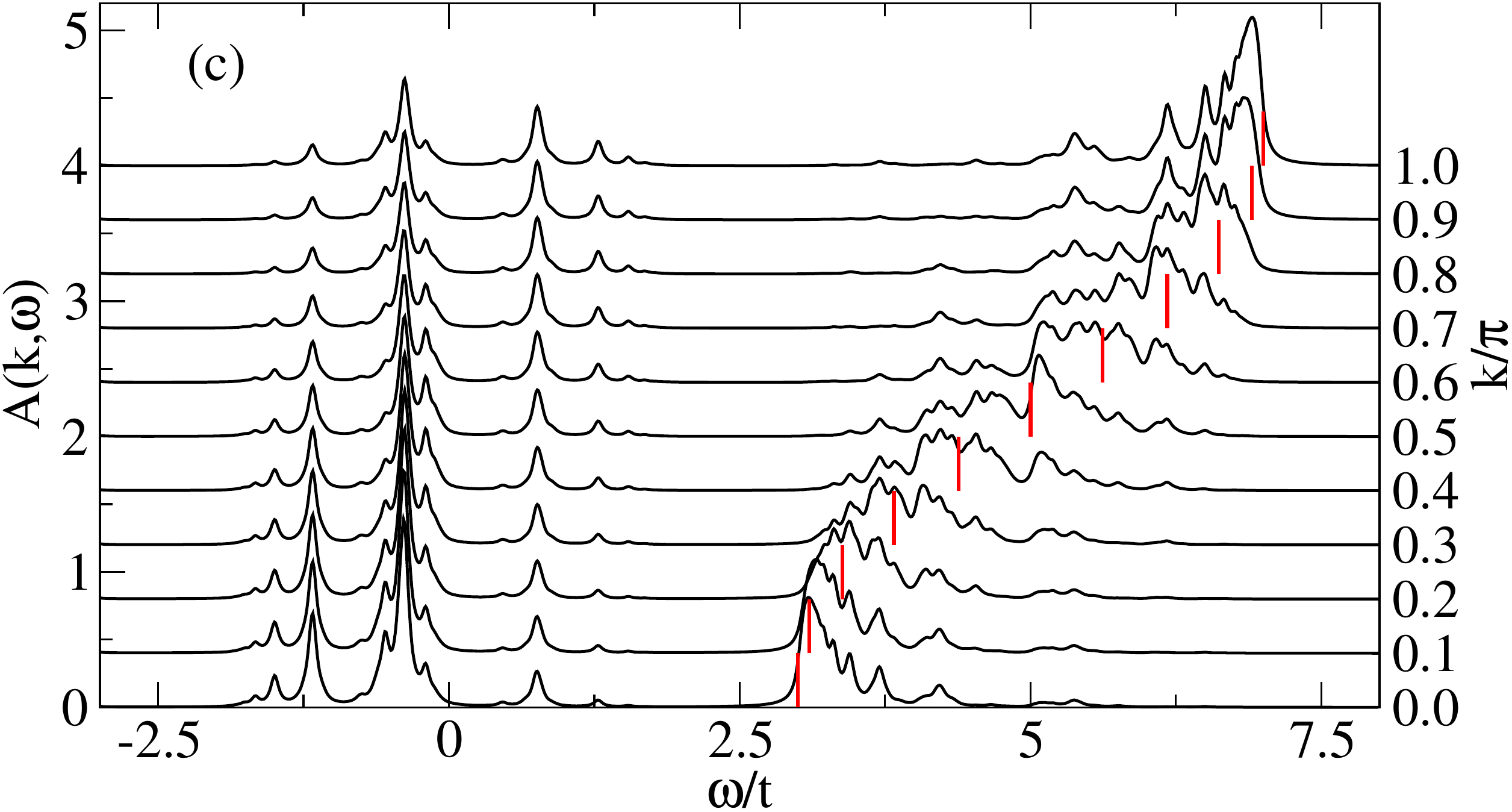}
   \caption{(color online) $A(k,\omega)$ for the three models 
with FM background at $\beta J=0.5$. Other parameters are as in Fig. \ref{fig2}. The
dispersionless low energy, low weight part of the spectrum of Model
III is not shown. Red, vertical lines indicate the location of the
$T=0$ quasiparticle peaks.}
  \label{fig3}
\end{figure}

With increasing $T$, both peaks broaden considerably on their
higher-energy side, and many resonances become visible. As
demonstrated in Ref. \onlinecite{Ising-finite-T} for Model II, these
resonances are due to temporal trapping of the carrier inside small
magnetic domains that are thermally generated at higher $T$. The
presence of these domains also explains the decreasing difference
between the $h=0$ and $h=-0.1t$ curves at higher $T$. For $\beta
J=0.5$ both curves are shown in the main panels (the finite $h$ curve
is shaded in). Indeed, the resonances appear in the same places and
with equal weight in both curves, the only difference being a small
spectral weight transfer from $J_0-2t$ to $-J_0-2t$, {\em i.e.} from
the FM ground-state disfavored by $h<0$ to the one favored by it. The
weight for the former is no longer zero like for $T\rightarrow 0$,
showing that at higher-$T$ the carrier is increasingly more likely to
explore longer domains of spin-up local moments.

The main difference between Models I and II is that the latter also
has a third finite-$T$ continuum, centered 
around $\omega=0$. It corresponds to injecting the carrier in small
AFM domains, where its exchange energy vanishes because it sits
between a spin-up and a spin-down local moment. Such energy
differences are not possible in Model II, where the carrier interacts
with a single moment so its exchange energy is $\pm J_0$.

However, if one is interested in the low-energy behavior, Models I and
II are equivalent because their low-energy continua have similar
origins and evolve similarly with $T$. This is true in the whole
Brillouin zone (BZ), as can be seen from comparing panels (a) and (b)
of Fig. \ref{fig3}.

The finite-$T$ evolution of the spectral weight of Model III is very
different. Consider first the $k=0$ case, shown in Fig. \ref{fig2}(c).
The $T=0$ peak at $2J-2t$ (marked by the vertical line)
evolves with  $T$ very 
similarly to the low-energy peaks of the other two models, broadening
on its high-energy side and again displaying resonances due to
temporal trapping inside small domains. The $k$ evolution of this
feature, shown in Fig. \ref{fig3}(c), is also very similar to the
low-energy continua of the other two models.

\begin{figure}[t]
  \centering 
  \includegraphics[width=0.6\columnwidth]{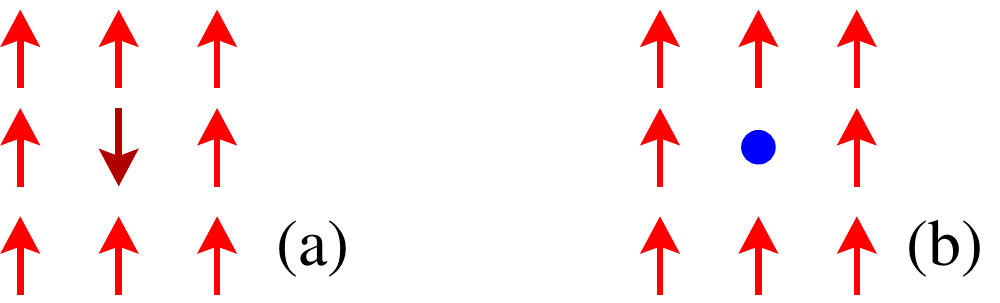}
  \caption{(color online) When doping ``removes'' a thermally excited
    spin-down, the energy variation upon doping is $E_b - E_a = -2dJ +
    \epsilon(k)$ and lies (at least partially) below the $T=0$
    quasiparticle ground-state energy of $2dJ-2dt$.}
  \label{fig4}
\end{figure}

However, for Model III this continuum is {\em not} the low-energy
feature. Instead, in 1d there are three lower-energy continua centered at
$0, -2J$ and $-4J$, all of which are due to injection of the carrier
into specific, thermally excited configurations of the
background. For example, consider the $-2dJ$ continuum which also
appears in dimensions $d>1$. As sketched in Fig. \ref{fig4}, it
corresponds to the carrier being paired with a thermally excited
spin. This lowers the exchange energy by
$2dJ$, as $2d$ AFM bonds are broken. In contrast, $T=0$ doping
always leads to loss of exchange energy, because only FM bonds can be 
broken. This is why in Model III it can cost less energy to dope from a thermally excited state rather than the ground-state, and therefore why
its finite-$T$ low-energy properties are not controlled by the $T=0$
quasiparticle. 

The weight of these low-energy continua is very small, see inset of
Fig. \ref{fig2}(c), because they are controlled by thermal
activation. For example, in the limit $T \rightarrow0$ the spectral
weight of the continuum centered at $-2dJ$ can be calculated to first
order, as was shown in Refs \onlinecite{low-T, Ising-finite-T}, by expanding Eq. (\ref{eq:GF-FT}) in powers of $\exponential{-\beta 4dJ}$. The lowest order terms correspond to the two FM ground states which have all spins aligned, $|\{\uparrow \}\rangle$ and $|\{ \downarrow \} \rangle$, respectively. The first order terms are given by states with a single flipped spin and are denoted by $|\{\uparrow, \sigma_m=\downarrow \} \rangle$ and $|\{\downarrow, \sigma_m=\uparrow \} \rangle$, where $m$ indicates the location of the flipped spin. Since the flipped spin can be anywhere in the system there are $N$ of these states for each ground state configuration. For simplicity we assume that the spin of the extra carrier is unpolarized, then it suffices to consider only $|\{\uparrow \} \rangle$ and $|\{\uparrow, \sigma_m = \downarrow \} \rangle$, the contribution from the other ground state will be exactly the same. Considering only these states in the trace of Eq. (\ref{eq:GF-FT}) we obtain:
\begin{align}
  G(\mathbf{k},\omega) = &\frac{1}{Z'} [\mathcal{G}^{(0)}(\mathbf{k},\omega) + \exponential{-\beta 4 dJ }\mathcal{G}^{(1)}(\mathbf{k},\omega)] \nonumber \\ &+ \mathcal{O}((\exponential{-\beta 4 d J})^2), 
\end{align}
where
\begin{align}
\mathcal{G}^{(0)}(\mathbf{k},\omega) = &[\omega - \epsilon_\mathbf{k}-2dJ + i \eta]^{-1} \\
\mathcal{G}^{(1)}(\mathbf{k},\omega) = & \sum_{n,\sigma} \exponential{i\mathbf{k}\mathbf{R}_n} \sum_m g_{0,n}(\omega,\{\uparrow,\sigma_m=\downarrow\}) 
\label{eq:G(1)}\\
Z' = &\frac{Z}{\exponential{-\beta E_{\text{FM}}}} = (1+N \exponential{-\beta 4dJ}+\dots)
\end{align}
Note that $\mathcal{G}^{(0)}(\mathbf{k},\omega)$ is identical to the $T=0$ solution. 

To evaluate $\mathcal{G}^{(1)}(\mathbf{k},\omega)$ we need to treat the case $m=0$, separately. In this case the extra carrier removes the flipped spin. This results in the breaking of $2d$ AFM bonds and therefore an energy gain of $2dJ$. Furthermore as pointed out above only $g_{0,0}(\omega,\{\uparrow,\sigma_0=\downarrow\})$ contributes to the sum since the extra carrier was injected into a domain of length 1. Since the flipped spin was removed and all the remaining spins are aligned it is easy o calculate $g_{0,0}(\omega,\{\uparrow,\sigma_0=\downarrow\})$ which in the limit $N\rightarrow \infty$ becomes
\begin{align}
 g_{0,0}(\omega,\{\uparrow,\sigma_0=\downarrow\}) = \int \frac{d \mathbf{q}}{(2 \pi)^d} \frac{1}{\omega -\epsilon_{\mathbf{q}} +2dJ +i\eta}, 
\label{eq:g0-integral}
\end{align}
{\em i.e.} a continuum of states centered at $\omega=-2dJ$.

We are now left with calculating the remaining contributions to $\mathcal{G}^{(1)}(\mathbf{k},\omega)$ for which $m \neq 0$. This is not a trivial problem, but since there is only one flipped spin in the system and we are summing over all ``hole'' locations, we can approximate $g_{0,n}(\omega,\{\uparrow,\sigma_m=\downarrow\}) \approx g_{0,n}(\omega,\{\uparrow \})$. In doing so we neglect that the energy is lowered when the ``hole'' is adjacent to the flipped spin $\sigma_m$. Reinserting into Eq. (\ref{eq:G(1)}) we obtain
\begin{align}
  \mathcal{G}^{(1)}(\mathbf{k},\omega) \approx g_{0,0}(\omega, \{\uparrow,\sigma_0=\downarrow \}) + (N-1)\mathcal{G}^{(0)}(\mathbf{k},\omega),
\end{align}
where the factor $N-1$ in front of $\mathcal{G}^{(0)}$ is due to the sum over $m$. Note that this factor ensures that the $Z'$ in the Eq. for $G(\mathbf{k},\omega)$ is approximately canceled. Similarly one expects contributions from states with two or more well-separated flipped spins to cancel the $Z'$ in front of $g_{0,0}(\omega, \{\uparrow,\sigma_0=\downarrow \})$ \cite{low-T, Ising-finite-T}. Consequently the low-$T$ expansion of the Green's function gives:
\begin{align}
  G(\mathbf{k},\omega) \approx \mathcal{G}^{(0)}(\mathbf{k},\omega) + \exponential{-\beta 4dJ} g_{0,0}(\omega, \{\uparrow,\sigma_0=\downarrow \}).
  \label{Gapprox}
\end{align}
{\em i.e.} the spectral weight below the $T=0$ quasiparticle which is given by $g_{0,0}(\omega, \{\uparrow,\sigma_0=\downarrow \})$ vanishes like the probability $\exponential{-\beta 4 d J}$ to find a flipped spin.

\begin{figure}[t]
  \centering 
  \includegraphics[angle=0, width=\columnwidth]{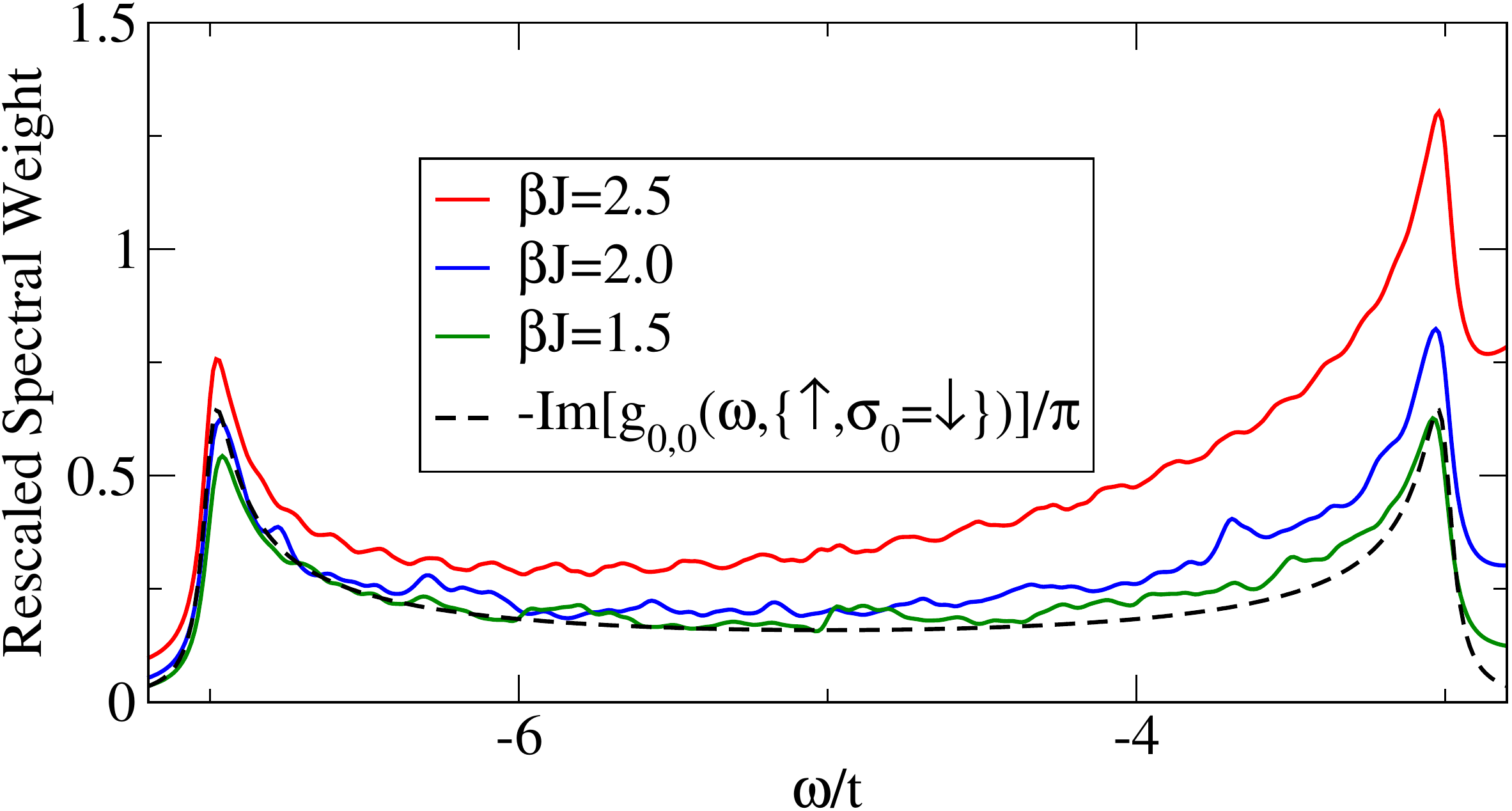}
  \caption{(color online) Rescaled spectral weight $\exponential{\beta 4 J}[A(0,\omega)-A^{(0)}(0,\omega)]$ in the region  of the  continuum centred at $-2J$, for Model III with FM background and different values of $\beta$. For comparison the dashed, black line shows $-\text{Im}[g_{0,0}(\omega,\{ \uparrow, \sigma_0 = \downarrow \})]/\pi$ calculated with Eq. (\ref{eq:g0-integral}). Other parameters are $J/t=2.5$ and $\eta/t=0.04$.}
  \label{fig5}
\end{figure}

To verify this behavior we show in Fig. \ref{fig5} the rescaled spectral weight $\exponential{\beta 4 J} [A(0,\omega)-A^{(0)}(0,\omega)]$, where $A^{(0)}(k,\omega) = \delta(\omega - \epsilon_k - 2 J) /\pi$ is the $T=0$ quasiparticle peak. From Eq. (\ref{Gapprox}) it is clear that at sufficiently low $T$ the resulting curves should equal $-\text{Im} [g_{0,0}(\omega,\{\uparrow,\sigma_0 = \downarrow\})]/\pi$, which is shown by the dashed, black line in Fig. \ref{fig5}. Indeed we find that the three curves in Fig. \ref{fig5} for $\beta J = 1.5$, $2.0$ and $2.5$, respectively, collapse onto each other and onto the curve for $-\text{Im} [g_{0,0}(\omega,\{\uparrow,\sigma_0 = \downarrow\})]/\pi$. Close to the upper edge of the continuum the agreement starts to falter. This is because in addition to the $T=0$ peak there are other peaks in the spectral weight (see Fig. \ref{fig2}) whose tails contribute to the $-2J$ continuum and are not subtracted. Multiplying with $\exponential{\beta 4J}$ amplifies these tails. Similarly the oscillating features in Fig. \ref{fig5} are numerical artefacts which are amplified by the factor $\exponential{\beta 4J}$. 

Similar calculations can be performed for the other low-energy
features. All their spectral weights vanish as $T\rightarrow 0$
because they all originate from doping the carrier into a thermally
excited environment, which become less and less likely to occur in
this limit.

\begin{figure}[t]
  \centering 
  \includegraphics[angle=0, width=\columnwidth]{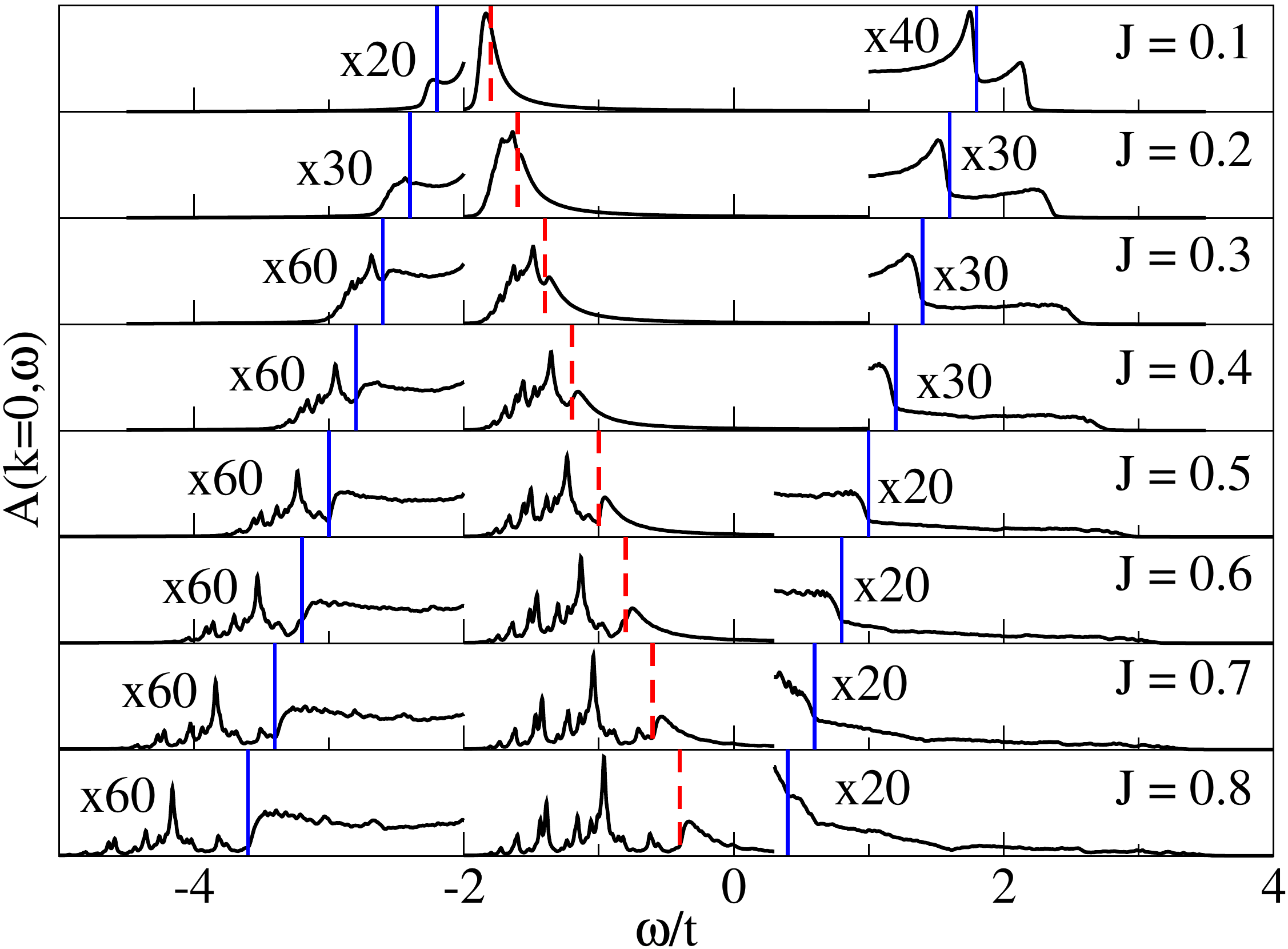}
  \caption{(color online) $A(k=0,\omega)$ for Model III with FM background, for different
    values of  $J$ at a temperature $\beta J =0.5$. The dashed red lines
    show the location of the $T=0$ quasiparticle peak. Full blue
    lines mark the energies $-2J \pm 2t$. Parts of the spectra have
    been rescaled for better visibility.}
  \label{fig6}
\end{figure}

The finite-$T$ behavior of Model III is thus {\em qualitatively}
different from that of Models I and II. For the latter, the $T=0$
quasiparticle peak also marks the lowest energy for electron-addition
at any finite temperature, whereas for Model III we observe the
appearance of electron-addition states well below the $T=0$
quasiparticle peak. Their spectral weight vanishes as $T\rightarrow
0$, which is very reminiscent of pseudogap behavior and offers a
simple and general scenario for how it can be generated. These
low-energy states vanish from the spectrum as the temperature is
lowered not because a gap opens and/or the electronic properties are
somehow changed, but simply because these states describe doping into
thermally excited local configurations, and the probability for the
doped carrier to encounter them vanishes as $T\rightarrow 0$.

As should be clear from these arguments, the appearance of these
low-energy continua is not a consequence of the large $J/t$ values
used so far for Model III. Indeed, Fig. \ref{fig6} shows that
similar behavior is observed for smaller $J$ values (parts of these
spectral weights have been rescaled for better visibility). With
decreasing $J$ the different continua overlap, but shoulders marking
some of their edges are still clearly visible and marked by dashed
lines. In all cases, at finite $T$ spectral weight appears below the
$T=0$ quasiparticle peak, marked by the full line.

It should also be clear that this phenomenology is not restricted to
FM backgrounds, either: one can easily think of excited configurations
in an AFM background whose exchange energy would be lowered through
doping, in a $t$-$J$ model similar to Model III. We have verified
numerically that at finite-$T$, features lying below the corresponding
$T=0$ quasiparticle peak indeed appear in the spectral weight of AFM
chains. These results are presented below. This
phenomenology is therefore quite general.

\subsection{AFM Results}
\label{sec:AFM_results}

\begin{figure}[t]
  \centering \includegraphics[angle=0,
    width=\columnwidth]{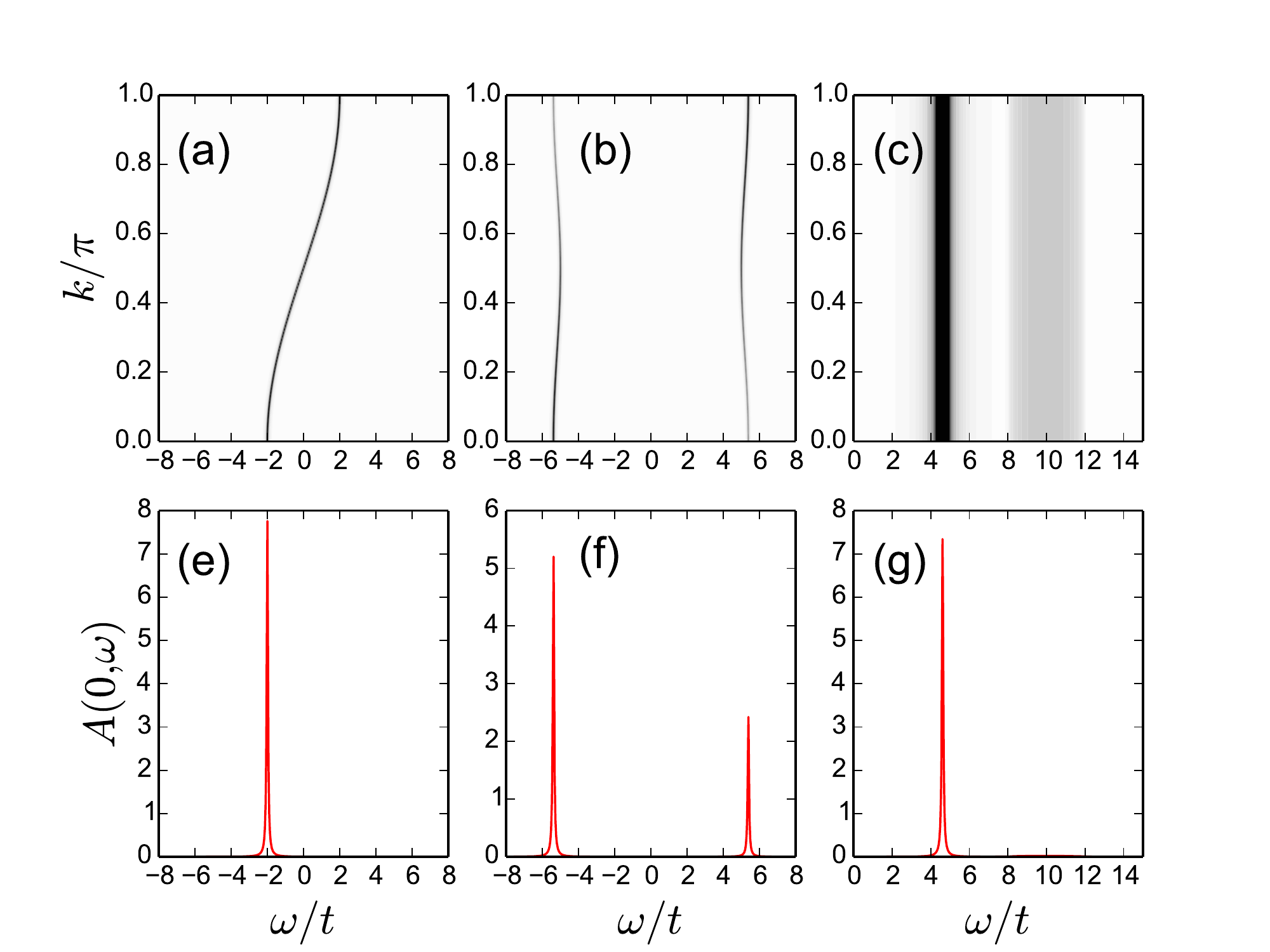}
  \caption{(color online) $T=0$, AFM solutions. Top panels: contour
    plots of $A(k,\omega)$. Bottom panels: Cross sections at $k=0$.
    (a) and (d) Model I with $J_0/t=5$, $|J|/t=0.5$; (b) and (e) Model
    II with $J_0/t=5$, $|J|/t=0.5$; (c) and (f) Model III with
    $|J|/t=2.5$. To improve visibility of the continuum a hard cutoff
    at $A(k,\omega)=0.1$ was used for the Model III contour plot. In
    all cases $\eta/t=0.04$.}
  \label{fig7}
\end{figure}

Just like in the FM case, there are also two  ground states
of the undoped AFM Ising chain: either the odd or the even lattice
hosts the up spins. Of course, both AFM ground states yield the same
quasiparticle properties. However, the quasiparticles that result when a
carrier is injected in the three models are different even at
$T=0$, for the AFM backgrounds. This is shown in Fig. \ref{fig7},
where contour plots of the $T=0$ spectral weight $A(k,\omega)$, and
cross sections at $k=0$, are shown.

\begin{figure}[t]
  \centering
  \includegraphics[width=\columnwidth]{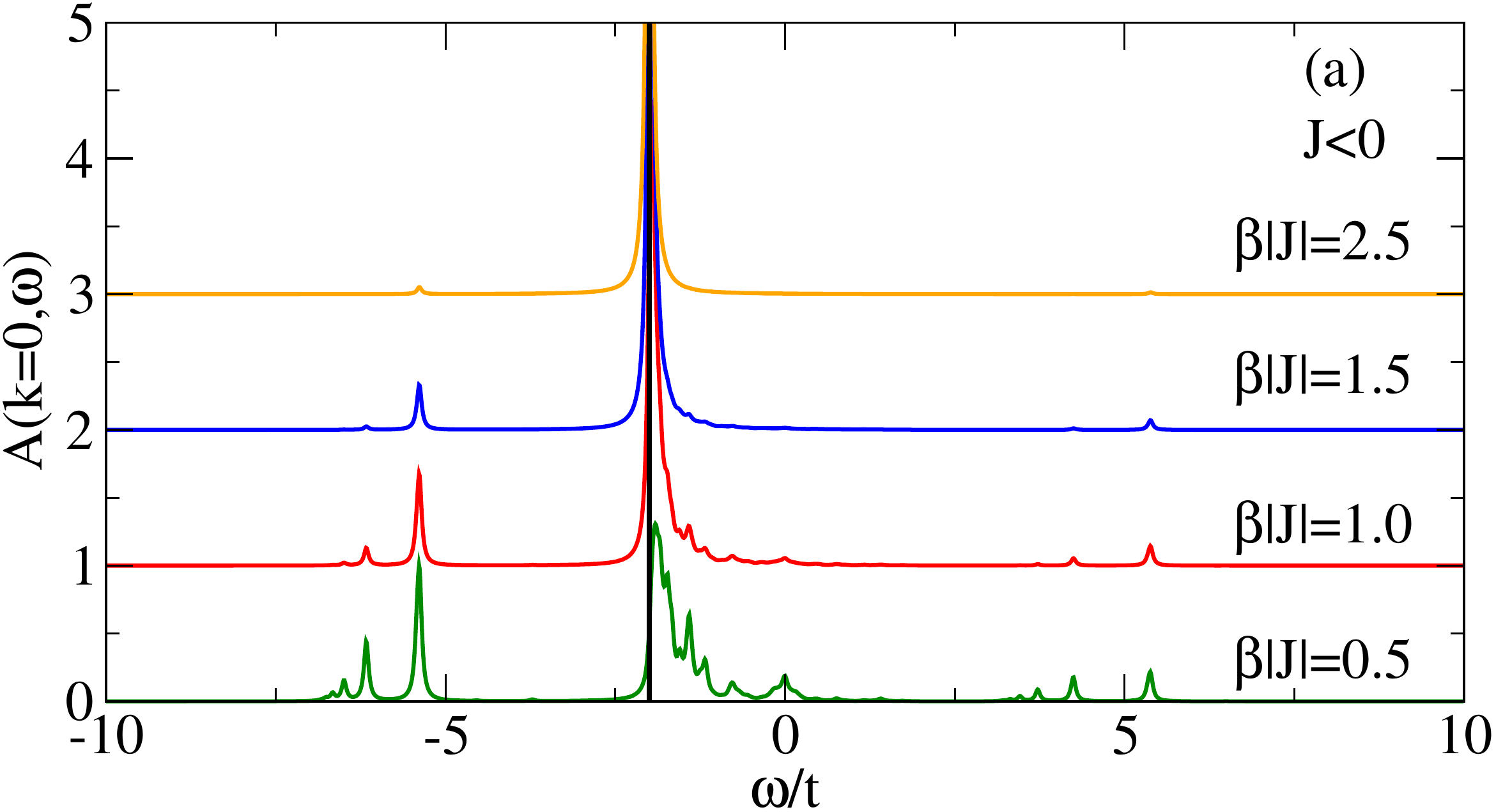}
  \includegraphics[width=\columnwidth]{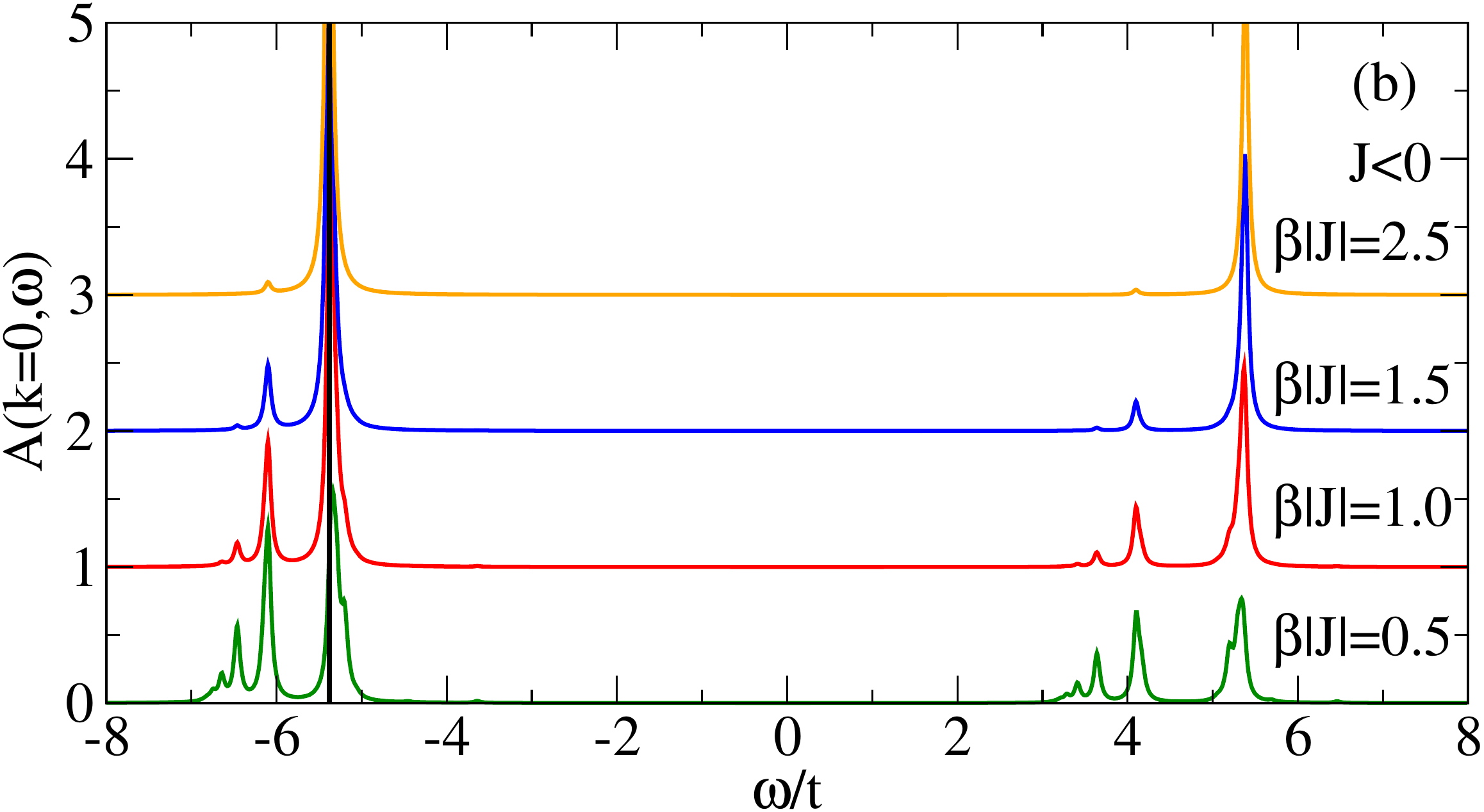}
  \includegraphics[width=\columnwidth]{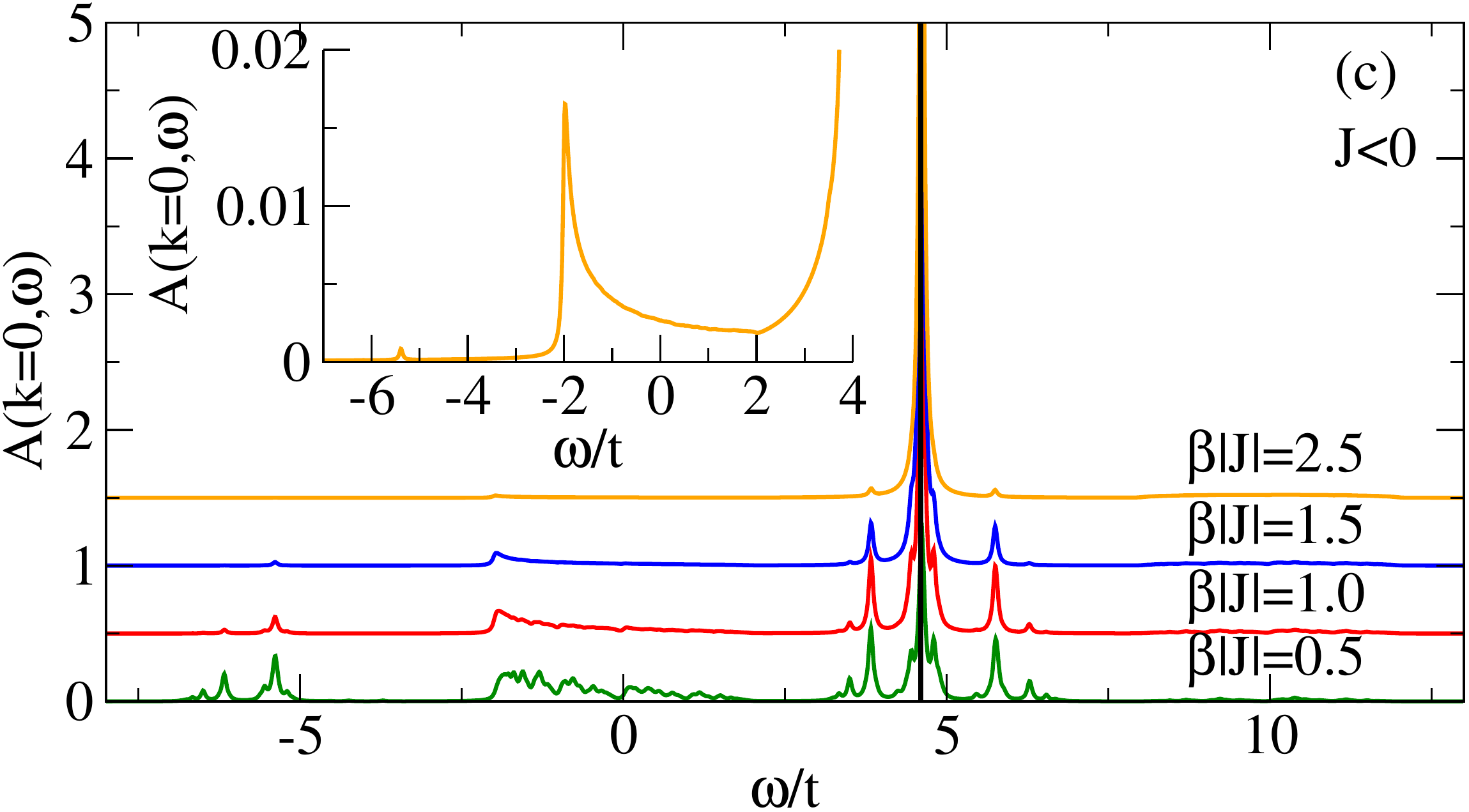}
  \caption{(color online) AFM Spectral weight at $k=0$ for three different
temperatures for (a) Model I with $J_0/t=5, |J|/t=0.5$; (b) Model II
with $J_0/t=5, |J|/t=0.5$; (c) Model III with $|J|/t=2.5$, the inset shows spectral weight below the $T=0$ quasiparticle peak for $\beta |J|=2.5$. In all cases,the broadening is $\eta/t =0.04$. The vertical lines show the energy of the $T=0$ quasiparticle peak.} 
  \label{fig8}
\end{figure}

For Model I, the energy shifts due to the Ising exchange with the
spins to the left and right of the extra electron exactly cancel out
and the quasiparticle behaves like a free electron with dispersion
$\epsilon(k)$. For Model II, interaction with the AFM background opens
a gap in the quasiparticle spectrum and halves its BZ. The upper and
lower bands have dispersion $\pm \sqrt{J_0^2 + \epsilon^2(k)}$,
respectively. For Model III, the $T=0$ spectral function is
independent of $k$ and has a coherent quasiparticle peak at $\omega =
4|J|-2\sqrt{J^2+t^2}$ and a continuum for $4|J|-2t< \omega < 4|J|+2t$.
The quasiparticle peak corresponds to a bound state with the extra
electron confined at its injection site. Propagation of the extra
electron along the chain reshuffles the Ising spins and gives rise to
the continuum centered at $4|J|$. One can therefore think of this
continuum as the electron+magnon continuum. Mathematically, the $k$-independence
follows directly from the fact that for Model III,
$g_{0,n}(\omega,\{\sigma\})=0$ when $n>0$, if $\{\sigma\}$ is
the AFM ground state.

The finite-$T$ spectral functions for $k=0$ are shown in Fig.
\ref{fig8}. For all three models the peaks broaden and spectral weight
appears below, as well as above the $T=0$ quasiparticle peak. This is
in contrast to the FM case, where in Models I and II, at $k=0$,
spectral weight appears only above the $T=0$ quasiparticle peak. For
Model I the energy difference between the low-energy states and the
$T=0$ peak is controlled by $J_0$, whereas for Model II it is of the
order of $t$. Just as for the FM case, these features can be linked to
small domains which temporarily trap the carrier
\cite{Ising-finite-T}. As the temperature increases more weight is
transferred to these low-energy features, and the low-energy behavior
of Model II starts to resemble that of Model I even though they have
different $T=0$ quasiparticles.

For Model III, at finite-$T$ new features appear, centered at $-2|J|$
and $0$. Just as for the FM case, they are due to the injection of the
extra electron into specific, excited, local configurations of the
chain. If the electron is injected into a small FM domain embedded in
an otherwise AFM ordered background, the energy is lowered by $2d|J|$.
As long as the electron stays within the FM domain reshuffling of the
spins does not result in a further change in energy. This explains the
appearance of spectral weight at $\omega \sim -2d|J|$. This is true
for any dimension $d$ and consequently the appearance of spectral
weight at $-2d|J|$ is a generic feature of Model III. Coming back to
the specific case of $d=1$, if the electron leaves the FM domain
reshuffling of the spins recreates an FM bond and destroys one of the
AFM bonds. The total change in energy (injection and reshuffling) is
therefore zero, explaining the continuum centered around $\omega \sim
0$. Besides the appearance of these new, low-energy continua,
resonances appear close to the $T=0$ quasiparticle peak and within the
high-energy continuum (not visible in Fig. \ref{fig8} due to the
scale). They are likely caused by injection of the electron into an
AFM domain and subsequent scattering off domain walls which can only
exist at finite-$T$.

\section{Conclusions}
\label{sec:conclusions}

In this work, we identified models that have identical $T=0$ low-energy
quasiparticles (for couplings favoring a FM background), and yet exhibit very different low-energy behavior at
finite $T$, proving that the former condition does not automatically
guarantee the latter.

In particular, the finite-$T$ behavior in Model III is controlled by
rare events, where the carrier is injected into certain magnetic
configurations created by thermal fluctuations. Their energies are
higher than that of the undoped ground-state, however the spectral
weight measures the {\em change} in energy upon carrier addition (or
removal), and this may be lower at finite-$T$ than at $T=0$. This is
the case for Model III because here doping removes a magnetic moment
from the background while its motion reshuffles the other ones. It is
not the case for Models I and II where the carrier can do neither of
these things. This difference is irrelevant at $T=0$ because of the
simple nature of the undoped FM ground-state, but becomes relevant at
finite-$T$.

We showed that such transfer of finite-$T$ spectral weight well
below the $T=0$ quasiparticle peak is independent of the size of the
magnetic coupling $J$ and occurs for both FM and AFM coupling.
Furthermore, we provided arguments that  this behavior is
expected to occur in any dimension.

While far from being a comprehensive study, these results clearly
demonstrate that the appearance of finite-$T$ spectral weight well
below the quasiparticle peak, due to the injection of the carrier
into a thermally excited local environment making it behave very
unlike the $T=0$ quasiparticle, is a rather generic feature for
$t$-$J$ like models. The weight of these finite-$T$, low-energy
features must vanish when $T\rightarrow 0$ because the probability for
such excited environments to occur vanishes, therefore these models
exhibit generic pseudogap behavior.

\begin{acknowledgments}
This work was supported by NSERC, QMI and the UBC 4YF (M.M.M.).
\end{acknowledgments}


\begin{thebibliography}{12}%
\makeatletter
\providecommand \@ifxundefined [1]{%
 \@ifx{#1\undefined}
}%
\providecommand \@ifnum [1]{%
 \ifnum #1\expandafter \@firstoftwo
 \else \expandafter \@secondoftwo
 \fi
}%
\providecommand \@ifx [1]{%
 \ifx #1\expandafter \@firstoftwo
 \else \expandafter \@secondoftwo
 \fi
}%
\providecommand \natexlab [1]{#1}%
\providecommand \enquote  [1]{``#1''}%
\providecommand \bibnamefont  [1]{#1}%
\providecommand \bibfnamefont [1]{#1}%
\providecommand \citenamefont [1]{#1}%
\providecommand \href@noop [0]{\@secondoftwo}%
\providecommand \href [0]{\begingroup \@sanitize@url \@href}%
\providecommand \@href[1]{\@@startlink{#1}\@@href}%
\providecommand \@@href[1]{\endgroup#1\@@endlink}%
\providecommand \@sanitize@url [0]{\catcode `\\12\catcode `\$12\catcode
  `\&12\catcode `\#12\catcode `\^12\catcode `\_12\catcode `\%12\relax}%
\providecommand \@@startlink[1]{}%
\providecommand \@@endlink[0]{}%
\providecommand \url  [0]{\begingroup\@sanitize@url \@url }%
\providecommand \@url [1]{\endgroup\@href {#1}{\urlprefix }}%
\providecommand \urlprefix  [0]{URL }%
\providecommand \Eprint [0]{\href }%
\providecommand \doibase [0]{http://dx.doi.org/}%
\providecommand \selectlanguage [0]{\@gobble}%
\providecommand \bibinfo  [0]{\@secondoftwo}%
\providecommand \bibfield  [0]{\@secondoftwo}%
\providecommand \translation [1]{[#1]}%
\providecommand \BibitemOpen [0]{}%
\providecommand \bibitemStop [0]{}%
\providecommand \bibitemNoStop [0]{.\EOS\space}%
\providecommand \EOS [0]{\spacefactor3000\relax}%
\providecommand \BibitemShut  [1]{\csname bibitem#1\endcsname}%
\let\auto@bib@innerbib\@empty
\bibitem [{\citenamefont {Box}\ and\ \citenamefont {Draper}()}]{best}%
  \BibitemOpen
  \bibfield  {author} {\bibinfo {author} {\bibfnamefont {G.~E.~P.}\
  \bibnamefont {Box}}\ and\ \bibinfo {author} {\bibfnamefont {N.~R.}\
  \bibnamefont {Draper}},\ }\href@noop {} {\emph {\bibinfo {title} {Empirical
  Model-Building and Response Surfaces}}}\ (\bibinfo  {publisher} {Wiley,
  1987})\BibitemShut {NoStop}%
\bibitem [{\citenamefont {Emery}(1987)}]{Emery}%
  \BibitemOpen
  \bibfield  {author} {\bibinfo {author} {\bibfnamefont {V.~J.}\ \bibnamefont
  {Emery}},\ }\href {\doibase 10.1103/PhysRevLett.58.2794} {\bibfield
  {journal} {\bibinfo  {journal} {Phys. Rev. Lett.}\ }\textbf {\bibinfo
  {volume} {58}},\ \bibinfo {pages} {2794} (\bibinfo {year}
  {1987})}\BibitemShut {NoStop}%
\bibitem [{\citenamefont {Lee}\ \emph {et~al.}(2006)\citenamefont {Lee},
  \citenamefont {Nagaosa},\ and\ \citenamefont {Wen}}]{Lee}%
  \BibitemOpen
  \bibfield  {author} {\bibinfo {author} {\bibfnamefont {P.~A.}\ \bibnamefont
  {Lee}}, \bibinfo {author} {\bibfnamefont {N.}~\bibnamefont {Nagaosa}}, \ and\
  \bibinfo {author} {\bibfnamefont {X.-G.}\ \bibnamefont {Wen}},\ }\href
  {\doibase 10.1103/RevModPhys.78.17} {\bibfield  {journal} {\bibinfo
  {journal} {Rev. Mod. Phys.}\ }\textbf {\bibinfo {volume} {78}},\ \bibinfo
  {pages} {17} (\bibinfo {year} {2006})}\BibitemShut {NoStop}%
\bibitem [{\citenamefont {Dagotto}(1994)}]{Dagotto}%
  \BibitemOpen
  \bibfield  {author} {\bibinfo {author} {\bibfnamefont {E.}~\bibnamefont
  {Dagotto}},\ }\href {\doibase 10.1103/RevModPhys.66.763} {\bibfield
  {journal} {\bibinfo  {journal} {Rev. Mod. Phys.}\ }\textbf {\bibinfo {volume}
  {66}},\ \bibinfo {pages} {763} (\bibinfo {year} {1994})}\BibitemShut
  {NoStop}%
\bibitem [{\citenamefont {Zhang}\ and\ \citenamefont
  {Rice}(1988)}]{Zhang-Rice}%
  \BibitemOpen
  \bibfield  {author} {\bibinfo {author} {\bibfnamefont {F.~C.}\ \bibnamefont
  {Zhang}}\ and\ \bibinfo {author} {\bibfnamefont {T.~M.}\ \bibnamefont
  {Rice}},\ }\href {\doibase 10.1103/PhysRevB.37.3759} {\bibfield  {journal}
  {\bibinfo  {journal} {Phys. Rev. B}\ }\textbf {\bibinfo {volume} {37}},\
  \bibinfo {pages} {3759} (\bibinfo {year} {1988})}\BibitemShut {NoStop}%
\bibitem [{\citenamefont {Anderson}(1987)}]{Anderson}%
  \BibitemOpen
  \bibfield  {author} {\bibinfo {author} {\bibfnamefont {P.~W.}\ \bibnamefont
  {Anderson}},\ }\href {\doibase 10.1126/science.235.4793.1196} {\bibfield
  {journal} {\bibinfo  {journal} {Science}\ }\textbf {\bibinfo {volume}
  {235}},\ \bibinfo {pages} {1196} (\bibinfo {year} {1987})}\BibitemShut
  {NoStop}%
\bibitem [{\citenamefont {Lau}\ \emph {et~al.}(2011)\citenamefont {Lau},
  \citenamefont {Berciu},\ and\ \citenamefont {Sawatzky}}]{Bayo}%
  \BibitemOpen
  \bibfield  {author} {\bibinfo {author} {\bibfnamefont {B.}~\bibnamefont
  {Lau}}, \bibinfo {author} {\bibfnamefont {M.}~\bibnamefont {Berciu}}, \ and\
  \bibinfo {author} {\bibfnamefont {G.~A.}\ \bibnamefont {Sawatzky}},\
  }\href@noop {} {\bibfield  {journal} {\bibinfo  {journal} {Phys. Rev. Lett.}\
  }\textbf {\bibinfo {volume} {106}},\ \bibinfo {pages} {036401} (\bibinfo
  {year} {2011})}\BibitemShut {NoStop}%
\bibitem [{\citenamefont {Ebrahimnejad}\ \emph {et~al.}(2014)\citenamefont
  {Ebrahimnejad}, \citenamefont {Sawatzky},\ and\ \citenamefont
  {Berciu}}]{Hadi1}%
  \BibitemOpen
  \bibfield  {author} {\bibinfo {author} {\bibfnamefont {H.}~\bibnamefont
  {Ebrahimnejad}}, \bibinfo {author} {\bibfnamefont {G.~A.}\ \bibnamefont
  {Sawatzky}}, \ and\ \bibinfo {author} {\bibfnamefont {M.}~\bibnamefont
  {Berciu}},\ }\href {\doibase 10.1038/nphys3130} {\bibfield  {journal}
  {\bibinfo  {journal} {Nature Phys.}\ }\textbf {\bibinfo {volume} {10}},\
  \bibinfo {pages} {951} (\bibinfo {year} {2014})}\BibitemShut {NoStop}%
\bibitem [{\citenamefont {Ebrahimnejad}\ \emph {et~al.}()\citenamefont
  {Ebrahimnejad}, \citenamefont {Sawatzky},\ and\ \citenamefont
  {Berciu}}]{Hadi2}%
  \BibitemOpen
  \bibfield  {author} {\bibinfo {author} {\bibfnamefont {H.}~\bibnamefont
  {Ebrahimnejad}}, \bibinfo {author} {\bibfnamefont {G.~A.}\ \bibnamefont
  {Sawatzky}}, \ and\ \bibinfo {author} {\bibfnamefont {M.}~\bibnamefont
  {Berciu}},\ }\href@noop {} {}\Eprint {http://arxiv.org/abs/arXiv:1505.04405}
  {arXiv:1505.04405} \BibitemShut {NoStop}%
\bibitem [{\citenamefont {M\"oller}\ and\ \citenamefont
  {Berciu}(2014)}]{Ising-finite-T}%
  \BibitemOpen
  \bibfield  {author} {\bibinfo {author} {\bibfnamefont {M.}~\bibnamefont
  {M\"oller}}\ and\ \bibinfo {author} {\bibfnamefont {M.}~\bibnamefont
  {Berciu}},\ }\href {\doibase 10.1103/PhysRevB.90.075145} {\bibfield
  {journal} {\bibinfo  {journal} {Phys. Rev. B}\ }\textbf {\bibinfo {volume}
  {90}},\ \bibinfo {pages} {075145} (\bibinfo {year} {2014})}\BibitemShut
  {NoStop}%
\bibitem [{\citenamefont {Mahan}(2000)}]{mahan}%
  \BibitemOpen
  \bibfield  {author} {\bibinfo {author} {\bibfnamefont {G.~D.}\ \bibnamefont
  {Mahan}},\ }\href@noop {} {\emph {\bibinfo {title} {Many-Particle Physics}}}\
  (\bibinfo  {publisher} {Springer},\ \bibinfo {year} {2000})\BibitemShut
  {NoStop}%
\bibitem [{\citenamefont {M\"oller}\ and\ \citenamefont
  {Berciu}(2013)}]{low-T}%
  \BibitemOpen
  \bibfield  {author} {\bibinfo {author} {\bibfnamefont {M.}~\bibnamefont
  {M\"oller}}\ and\ \bibinfo {author} {\bibfnamefont {M.}~\bibnamefont
  {Berciu}},\ }\href {\doibase 10.1103/PhysRevB.88.195111} {\bibfield
  {journal} {\bibinfo  {journal} {Phys. Rev. B}\ }\textbf {\bibinfo {volume}
  {88}},\ \bibinfo {pages} {195111} (\bibinfo {year} {2013})}\BibitemShut
  {NoStop}%
\end{thebibliography}

%

\end{document}